 \documentclass[preprint2]{aastex6}

\fullcollaborationName{The Friends of AASTeX Collaboration}

\begin{document}


\title{The Efficiency of Noble Gas Trapping in Astrophysical Environments}


\author{Fred J. Ciesla, Sebastiaan Krijt, Reika Yokochi}
\affil{Department of the Geophysical Sciences, University of Chicago, 5734 South Ellis Avenue, Chicago, IL, USA}
\author{Scott Sandford}
\affil{NASA Ames Research Center}







\begin{abstract}
Amorphous ice has long been invoked as a means for trapping extreme volatiles into solids, explaining the abundances of these species in comets and planetary atmospheres.
  Experiments have shown that such trapping is possible and have been used to estimate the abundances of each species in primitive ices after they formed.  However, these experiments have been carried out at deposition rates which exceed those expected in a molecular cloud or solar nebula by many orders of magnitude.  Here we develop a numerical model which reproduces the experimental results and apply it to those conditions expected in molecular clouds and protoplanetary disks.  We find that two regimes of ice trapping exist: \emph{burial trapping} where the ratio of trapped species to water in the ice reflects that same ratio in the gas and \emph{equilibrium trapping} where the ratio in the ice depends only on the partial pressure of the trapped species in the gas.  The boundary between these two regimes is set by both the temperature and rate of ice deposition.  Such effects must be accounted for when determining the source of trapped volatiles during planet formation.
\end{abstract}




\section{Introduction} \label{sec:intro}

The incorporation of noble gases into early planetesimals is of interest for many reasons.   The Galileo spacecraft found that the abundances of these elements relative to hydrogen, along with other volatile elements such as C and N, were elevated in Jupiter when compared to the solar nebula, suggesting that the giant planet's atmosphere was polluted by planetesimals with solar abundances of all elements except for H and He \citep{owen99}.  The presence of noble gases in terrestrial planet atmospheres has been suggested to have arisen at least in part due to accretion of comets over their histories \citep[e.g.][]{notesco03,notesco05,dauphas03,marty17}.  Such an idea is supported by recent observations from the Rosetta mission indicate that noble gases (Ar) are indeed present within the comet, showing that solid bodies did incorporate some amount of noble gases \citep{balsiger15}.

How these elements were incorporated into primitive bodies remains a mystery.  Noble gases have very low condensation temperatures  \citep[$<$50 K][]{gautier01,lodders03}, suggesting that these elements would not be contained within the solids present where planets formed. 
Water could serve as a carrier for these elements, trapping them within an icy matrix such that they would only be lost once higher temperatures were reached.  However, it is still debated how the water ice could incorporate the noble gases into its structure, with two primary methods currently being discussed.  In explaining Jupiter's atmospheric composition, \citet{owen99} proposed that the elements were trapped as amorphous water ice formed on solids, with guest species being surrounded and buried by water molecules freezing-out in cold environments.  Such trapping had been seen experimentally at very low temperatures \citep[$<$50 K, e.g.][]{barnun85, barnun88}, implying Jupiter  accreted solids that formed at the very distant edges of the Solar System.  \citet{gautier01} instead suggested that the noble gases were incorporated within crystalline water ice as clathrates, which were predicted to form at higher temperatures than the vaporization point of the guest molecules \citep{luninestevenson85,mousis16}.  

Amorphous ice and clathrates form under very different temperature and pressure conditions, thus identifying which form of ice the noble gases were originally locked away in will provide important insights into the history of water during the early stages of planet formation.
However, determining which of these forms was the dominant carrier for the noble gases (as well as other volatiles such as CO and N$_{2}$), we must understand the efficiency with which gases are incorporated into ice in the various environments that existed during the birth of our Solar System. Here we focus on the issue of trapping in amorphous ice.  While experimental studies have been used to estimate the conditions under which primitive ices may have formed \citep[e.g.][]{notesco05}, these experiments and those of  others \citep[e.g.][]{collings03,fayolle11,yokochi12}, were performed at very high deposition rates of water ice, which would imply freeze-out fluxes that are many orders of magnitude above those expected in astrophysical environments.  These experimental conditions are necessary in order to study the processes at work on laboratory time scales, however it is unclear how such results can be extrapolated to the very different conditions expected during planet formation.

Here we develop a mathematical model to understand how noble gases would be locked away in amorphous ices that formed in the ISM or outer solar nebula.  In the next section, we review the experimental work that has been done on noble gas trapping.  We then describe the three-phase model used to quantitatively investigate how water and guest species are exchanged between the gas, solid surface, and mantle of amorphous ice, fitting the model parameters to reproduce the experimental results and trends described by previous studies. We apply these models to various astrophysical environments in order to evaluate the efficiency of this process.  In interpreting these results, we consider what physical processes must be considered in future work, both theoretical and experimental, in order to evaluate the role that amorphous ice trapping played in setting the noble gas inventories of planetary bodies.

\section{Review of Noble Gas Trapping Experiments}

The trapping of volatile species by water ice has been studied by a number of authors using a variety of experimental setups \citep[e.g.][]{barnun88,sandfordall90,collings03,fayolle11,yokochi12}.  We focus on the studies by Bar-Nun and collaborators \citep[e.g.][]{barnun88,notesco99,notesco03,notesco05} in developing our numerical model as they have reported the largest collection of experimental results, and thus provide a set of numbers from a wide range of conditions that can be used to constrain key parameters in our model.  Further, it was these results that motivated \citet{owen99} to suggest that Jupiter's atmospheric composition may be explained by trapping in this manner.  Here we explain the conceptual framework for those experiments, and return to the results of other studies in our discussions further below.

In the experiments, a mix of water vapor and a
guest species (here we focus on Ar as a representative noble gas/volatile) flowed over a cold plate within an experimental chamber via cryopumping.  The cold plate was set at a given temperature and the rate of flow was controlled (remained constant) so that layers of ice measuring $\sim$0.1 $\mu$m thick were deposited  on timescales of minutes to days (deposition
rates of ice of ~10$^{-5}$-10$^{-1}$ $\mu$m/min).  After deposition, the experimental chamber was then pumped down to remove any remaining gas and the cold plate was heated at rates of $\sim$1-10 K/min. During heating, gas was continuously pumped away and its composition measured; the composition was assumed to reflect
what was sublimated from the ice at that time or temperature.

While Ar was found to be present in the chamber throughout the experiment, its abundance increased rapidly at particular instances during warming.  Immediately after heating began, the flux of Ar into the pump rose significantly above the background; this was interpreted as some amount of frozen Ar that was being from from the deposited solid.  That is, this Ar was not physically trapped within the ice but adsorbed onto the substrate, and thus was liberated to the gas phase due to direct thermal desorption.  The second, significant pulse of Ar came when temperatures reached $\sim$120K, when water molecules also began to desorb.  As this Ar was only released once water itself also began to be seen in the vapor, this was interpreted as Ar that was trapped within the amorphous ice--atoms that were unable to desorb due to a physical barrier provided by the water molecules.  

Again, Ar was found in the gas throughout the experiments.  Beyond vaporization, the amorphous water ice also would evolve physically, as molecules within the ice would diffuse or rearrange themselves when undergoing various phase transitions at temperatures $>$80 K. Release of vapor could occur as these physical changes take place, and have been reported in other experimental studies \citep{collings03,viti04}.  The total amount of Ar trapped in the Bar-Nun group's experiments was defined by that which was released after the first pulse of Ar had declined to background levels, which occurred once warming brought the sample to T$\sim$50 K \citep{notesco03,notesco05}.  

A few critical assumptions are made in interpreting the measurements from this study that are important to highlight.  The first is that all freezing-out and desorption of water and Ar occurred on the cold plate and nowhere else in the experimental chamber.  If materials were frozen out elsewhere, then it is possible that the Ar that is measured was not trapped, but instead originated from an unrelated region in the experimental chamber.  This would be particularly important as it was the cold-plate was warmed in the experiments and not the entire chamber; if temperature gradients were present, then it is possible that Ar that was frozen-out (not trapped) desorbed while the cold plate was at a higher temperature than the region where the Ar was actually released.  Finally, with Ar present throughout the warm-up phase, it is possible that the gas that was pumped out was not an immediate reflection of what was desorbed from the ice, meaning that the recorded compositions were influenced by what was vaporized at a lower temperature than when the data was recorded.  These issues were minimized by carrying out the experiments under high vacuum conditions and the use of line-of-sight methods to maximize measured desorption to specific sample region.  Nonetheless, these possibilities mean that the numbers from the experiments should be taken as upper limits on the amount of trapped Ar in the ice.  Here, we follow the authors by assuming that the measurements reflect the actual ice composition, but return to this issue in the discussion.

\section{Overview of the Three-Phase Model}

To understand how the composition of the ice evolved over time in these experiments, we adopt the basic three-phase model of \citet{hasegawaherbst93}.  This model tracks the exchange of species between the gas and solid phases, with the solid phase divided into the surface, which communicates with the gas, and the mantle, which does not.  We also extend the model by following \citet{fayolle11}, who established a means to allow for the exchange of species between the mantle and the surface.

Exchange between the solid phase and gas phase occurs through adsorption and desorption of the various species considered.  Adsorption occurs when a gaseous species collides with a solid substrate and sticks.  Desorption occurs when a molecule at the surface leaves the solid surface and returns to the gas because the thermal energy it attains in the solid is enough to overcome its binding energy to the substrate.  The mantle is formed when molecules or atoms freeze-out on top of already adsorbed species, burying them.   An atom or molecule is added to the surface from the mantle when it is exposed by desorption of an overlying species, or through diffusion (swapping positions with other species).

The rate of adsorption by a gaseous species and thus increase in the abundance of the surface species (per unit volume), $n_{i}^{s}$, is given by:
\begin{equation}
\frac{d n_{i}^{s}}{d t} = v_{th}^{i} n_{i}^{g}
\end{equation}
where $n_{i}^{g}$ is the number density of the molecule or atom in the gas,  $v_{th}^{i}$ is the thermal velocity of the gaseous species given by $v_{th}^{i}$=$\left( \frac{8 k T}{\pi m_{i}} \right)^{\frac{1}{2}}$, and we have assumed every collision of the gas molecule onto the surface leads to sticking.  Here the surface abundance is given by the number of molecules or atoms on the surface, $N_{s}^{i}$, times the number density of  surfaces (dust grains in the astrophysical setting), $n_{d}$; that is $n_{i}^{s} = N_{s}^{i} n_{d}$. 

The rate of desorption of a species from the surface to the gas is given by the Polanyi-Wigner Equation, which describes the thermal desorption of solid species from a substrate as it is warmed \citep[e.g.][]{bergin_pdchem11,smith16,chaabouni18}:
\begin{equation}
\frac{d n_{i}^{s}}{dt}=-\nu_{i} \exp{\left(-\frac{E_{i}}{T_{dust}}\right)} n_{i}^{s}
\end{equation}
where $n_{i}^{s}$ is the abundance of the molecule or atom at the surface of the solid, $E_{i}$ is the binding energy (in units of K as we take $E_{i}$=$E_{\mathrm{bind,i}}$/$k$), and $\nu$ is the vibrational frequency of the species in the potential well which keeps it bound to the surface and is of order $\nu \sim$10$^{12}$ s$^{-1}$  for the species of interest here \citep{biham01,fayolle11,bergin_pdchem11}.  While a single value for the binding energy is used here, the ice surface is likely heterogeneous, leading to a distribution of binding energies for a given species to the ice as demonstrated for CO binding to water ice \citep[e.g.][]{karssmeijer14}.  The single value used here should be considered the most probable binding energy of the whole distribution of binding energies \citep{smith16}.

\begin{figure}[!h]
\includegraphics[width=3.5in]{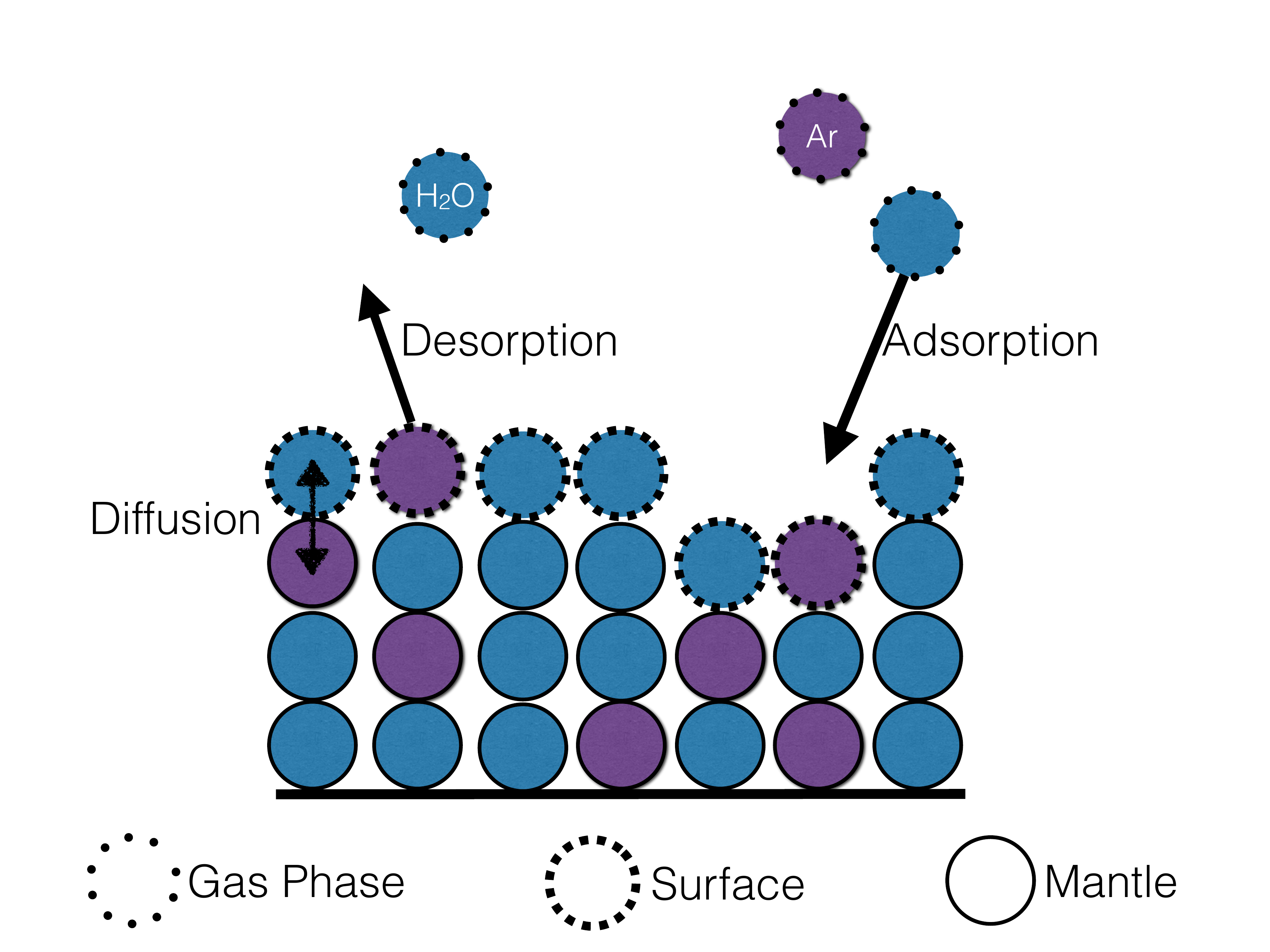}
\caption{Schematic diagram of the three-phase model described here.  We focus on the deposition and evolution of an H$_{2}$O-Ar mixture, where each species exists either in the gaseous, surface, or mantle phases.  Evolution among these phases occurs through adsorption, desorption, and diffusion. Model is based on that developed and described by \citet{fayolle11}.}
\end{figure}

The mantle forms when adsorbing species land on top of already adsorbed species.  As the landing point of adsorbed molecules is  random, the growth of the mantle during deposition is given by:
\begin{equation}
\frac{d n_{i}^{m}}{dt} = \alpha \frac{d n_{s}}{dt} \frac{n_{i}^{s}}{n_{s}}
\end{equation}
where $n_{i}^{m}$ is the abundance of the species in the mantle with $n_{i}^{m}=N_{m}^{i} n_{d}$.
Here $\frac{d n_{s}}{dt}$ is the total sum of adsorption rates minus the rates of desorption for all species considered, $\alpha$ is the fraction of surface sites occupied at the time of adsorption.  That is, a typical surface can host $\sim$10$^{15}$ sites/cm$^{2}$; if only a portion of those sites are occupied, an adsorbing species can either land on an already adsorbed species, moving that adsorbed species to the mantle, or it can fill a vacant surface site, thus not affecting the mantle at all.  Thus, $\alpha$ represents the probability of landing on an occupied surface site,  while (1-$\alpha$) represents the probability that an adsorbing species landed on bare substrate instead of previously adsorbed molecule or atom.  Note that the surface may represent multiple partially filled monolayers of adsorbed particles--it represents the collection of species which are directly exposed to the gas (see Figure 1).

When total desorption rates are greater than total adsorption rates, the mantle composition then evolves as surface species are removed, exposing buried species.  These exposed species now become part of the surface, and are no longer considered part of the mantle.  In this case, the exchange between the surface and mantle is given by:
 \begin{equation}
 \frac{d n_{i}^{s}}{dt} = \left[ \sum_{j} \nu e^{-\frac{E_{j}}{T}} n_{j}^{s} \right] \frac{n_{i}^{m}}{\sum_{j} n_{j}^{m}}
 \end{equation}

In the absence of any other processes, a guest atom or molecule would become trapped once another species adsorbs above it.  Release of the trapped species could only occur once all of the species that were adsorbed on top of it were desorbed. However, experiments have shown that volatiles incorporated into the mantles of deposited ice can be lost as well, suggesting that swapping of species or exchange between the mantle and surface can occur \citep[e.g.][]{oberg09,fayolle11}.  This was seen as greater amounts of the volatiles were lost from the ice at their respective desorption temperatures than could be explained from those present on the surface layer alone.  This indicates that some fraction of the volatiles in the ice are trapped as they are only lost along with the binding water molecules, while others remain mobile and make their way to desorb directly at much lower temperatures.  

 \textbf{The details of how migration of the volatile through the ice occurs remain uncertain, with some of it possibly occurring within pores and cracks in the ice.  However, trapping, and thus limitations on migration, is independent of the pores and cracks available, as species with different volatilities (binding energies) such as CO and CO$_{2}$ are released from ices at very different temperatures, which cannot be explained simply by physical restructuring or evolution of these pores and cracks \citep{fayolle11}.}

\textbf{ In the experimental studies, certain trends and relations were observed that must be reproduced in terms of the  loss of a non-trapped component of the volatile.   For example, it was found that the fraction of a trapped CO and CO$_{2}$ increased with increasing ice thickness \citep{fayolle11}. This is consistent with the findings of \citet{notesco05}, who found that higher trapped Ar/H$_{2}$O ratios were found in ice layers 5 $\mu$m thick when compared to those 0.1 $\mu$m thick.  Further, these experiments also showed that the fraction of the deposited volatile that was trapped increased as its relative abundance in the ice decreased.  In other words, for the same amount of water, higher amounts of the trapped volatile led to greater fractions being lost during heating instead of trapped.}

 To account for the loss of volatiles in a manner that is consistent with the experimental behaviors described above,
\citet{fayolle11} extended the three-phase model by allowing molecules or atoms in the mantle to migrate to the surface.   Such a model efficiently reproduced the desorption of species from a deposited CO$_{2}$+H$_{2}$O and CO+H$_{2}$O mixtures.    The rate of exchange between the ice surface and mantle is given by:
\begin{equation}
R_{i}^{\mathrm{diff}} = \frac{f_{i} \nu}{\sum_{j} n_{j}^{m}} e^{-\frac{E_{\mathrm{diff}}}{T}} 
\left[ n_{\mathrm{H_{2}O}}^{s} n_{i}^{m} - n_{i}^{s}  n_{\mathrm{H_{2}O}}^{m}  \right]
\end{equation}
Here $E_{\mathrm{diff}}$ is the energy of diffusion required for the species to swap locations with a surface water molecule.  This energy can be related to the binding energy of species of interest; for simplicity we follow \citet{garrod13} by assuming they are proportional and set the value for $E_{\mathrm{diff}}$=0.5$E_{i}$.  We discuss how various values of $E_{\mathrm{diff}}$ would affect the results of this study further below.  

Above, $f_{i}$ represents the fraction of the non-water mantle that is able to migrate to the surface of the ice.  That is, experiments suggest that only a portion of the mantle communicates with the surface to allow for migration of the volatile guest, and that this varies with the relative abundance of that volatile contained within the mantle during deposition.
Based on experimental studies of \citet{oberg09}, the form of $f_{i}$ is given by:
\begin{equation}\label{eq:frac}
f_{i} = 1 - \frac{ n_{i}^{m,ini} - c_{i} \left(x_{i}^{ini} \right)^{\beta}}{n_{i}^{m}}
\end{equation}
where $n_{i}^{m,ini}$ represents the initial abundance of the species of interest in the mantle after deposition and before heating and $x_{i}^{ini}$ is the initial mixing ratio of the species to water in the mantle, $x_{i}$=$n_{i}^{m,ini}$/$n_{\mathrm{H_{2}O}}^{ini}$.  The parameter, $c_{i}$, is the availability constant and describes the extent to which the mantle communicates with the surface.  That is, given that $f_{i}$ defines the fraction of the mantle species able to diffuse to the surface and be desorbed upon heating, (1-$f_{i}$) is the fraction that remains trapped within the water ice.  Thus knowing $c_{i}$, one can predict the amount of trapped guest molecules to remain in the ice:
\begin{equation}
n_{i}^{tr}=n_{i}^{m,ini} - c_{i} \left(x_{i}^{ini} \right)^{\beta}
\end{equation}
where we look at the instant immediately after deposition when $n_{i}^{m}$=$n_{i}^{m,ini}$.  What will occur upon heating is $c_{i} \left(x_{i}^{ini} \right)^{\beta}$ species will migrate to the surface of the ice and desorb as part of the frozen out species released initially upon warming as seen in the experiments described above.  The fraction of volatile that is trapped is given by:
\begin{equation}
f_{i}^{tr} = 1 - \frac{c_{i}}{n_{\mathrm{H_{2}O}}^{ini}}x_{i}^{\beta-1}
\end{equation}
Thus, provided $\beta >$1, as $x_{i}$ decreases, the fraction of the original volatile that is trapped increases, in agreement with experiments \citep{oberg09,fayolle11}.  Further, for the same conditions, as the ice gets thicker (as $n_{\mathrm{H_{2}O}}^{ini}$ increases) the fraction of trapped volatile also increases, again in line with experimental results described above.
Following \citet{oberg09}, we take $\beta$=2, but have also explored other values (1-5) and found that they have little effect on our conclusions below.  Further, we do not have to worry if the availability constant varies with ice thickness as we consider just ice layers 0.1 $\mu$m thick to be consistent with the experiments to which we are fitting model parameters.  

\textbf{The purpose of the availability constant is thus to set how the deposited volatile is distributed between the trapped component and the component the frozen component.  Thus it is a measure of how mobile a given species is.  Note that if $c_{i}$=0, then all of the species in the mantle are trapped, as no migration would occur.  As $c_{i}$ increases, the amount of trapped material decreases, as more and more of the guest species is able to migrate from the mantle to the surface.  Also, as $x_{i}$, the ratio of the volatile to water in the deposited ice increases, Equation (8) predicts that the trapped fraction decreases.  Further, for higher values of $n_{\mathrm{H_{2}O}}^{ini}$, or thicker ices, the fraction of trapped volatile increases.  Thus, this captures the behavior of the major effects observed in the experimental results described above.}

Because water is expected to be much more abundant than argon in the solar nebula and molecular clouds, the details of diffusion and migration of untrapped species is likely unimportant, unless only portions of the water budget are in the vapor phase, resulting in high Ar/H$_{2}$O ratios in the gas.  We return to this point in the Summary and Discussion section.


While the above approach is general enough for any species of interest, in the application of the model below, we continue to focus on Ar as a representative noble gas or volatile that could be trapped by the ice.  The only unknowns in this model are the binding energy of Ar to the substrate, $E_{\mathrm{Ar}}$, and the availability constant, $c_{\mathrm{Ar}}$.  As such, we vary these parameters in order to determine which provides the best match to the experimental results described above.

\section{Modeling Experimental Results}

In order to fit the parameters in our model to produce the results reported by \citet{notesco03, notesco05}, the experimental conditions must be translated to variables to be used in the equations described above.
 We divide the experiment into two parts: deposition and warming.  Deposition corresponds to the time when water and Ar are injected into the experimental chamber and freeze-out onto the gold-coated copper plate that, as reported by \citet{barnun87}, measured 5 cm $\times$2.5 cm for a total 12.5 cm$^{2}$ in area.  As a test in those experiments, a 2 $\mu$m thick layer of ice developed which was estimated to correspond to 10$^{19}$-10$^{20}$ H$_{2}$O molecules.  Assuming uniform thickness of the ice across the copper plate, and that species were only deposited on the plate and nowhere else in the apparatus, this corresponds to each molecule occupying a space of $\sim$2.5$\times$10$^{-22}$-2.5$\times$10$^{-23}$ cm$^{3}$, or a region with a linear lengthscale of $L \sim$3$\times$10$^{-8}$ to 6$\times$10$^{-8}$ cm.  The number of occupied sites in a monolayer (ML) per unit area is then 1/$L^{2}$ $\sim$ 0.3-1.3$\times$10$^{15}$ cm$^{-2}$.  Here, we will take $N_{s}=$10$^{15}$ cm$^{-2}$ ML$^{-1}$, which is consistent with astrochemical studies as described above \citep[e.g.][]{hollenbach09, bergin_pdchem11, fayolle11}.  In depositing a 2 $\mu$m thick layer of ice, this means that there were  $\sim$3200-6900 monolayers of ice in the sample.  Taking 5000 layers as typical, then these estimates give a total number of deposited molecules in the experiments of:
$
10^{15} \mathrm{~molecules ~cm^{-2} ~layer^{-1}}  \times 5000 \mathrm{~layers} \times 12.5 \mathrm{~cm^{2}} = 6.25 \times 10^{19} \mathrm{~molecules}
$, a value in line with the estimates reported by \citet{barnun87}.  We focus on a particular 
unit area (1 cm$^{2}$) suspended such that the number of surface sites available is $N_{s}$=10$^{15}$ species per layer, and that the volume density of particles, $n_{d}$=1 cm$^{-3}$.

Given these estimates, the flux of molecules onto the plate can be estimated.  The fluence needed to build a 1 $\mu$m thick layer of ice is 2500 ML $\times$10$^{15}$ sites cm$^{-2}$ ML$^{-1}$=2.5$\times$10$^{18}$ molecules cm$^{-2}$.  The deposition rates used in the experiments described above ranged from $DR$=10$^{-5}$-10$^{-1}$ $\mu$m min$^{-1}$, which would require adsorption fluxes of:
\begin{equation}
F_{dep} = 4.2 \times 10^{15} \left( \frac{DR}{0.1 \mathrm{ ~\mu m ~min^{-1}}} \right) ~\mathrm{ species ~cm^{-2} ~s^{-1}}
\end{equation}
This can be equated to the flux expected for gaseous species, $F_{i} \sim \frac{1}{4} n_{\mathrm{i}} v^{i}_{th}$.  Thus for a given temperature, we set the adsorption rate to the flux defined by the experimental deposition rate, with total H$_{2}$O and Ar fluxes summing to $F_{dep}$.  These are set as inputs into our model, with the gaseous number densities and temperatures held constant throughout the period of deposition (that is, as species freeze-out, we assume that they are replaced through the experimental apparatus to keep the flux constant) as gas was constantly replenished by the steady flow over the plate in the experiments.  Given these inputs, we calculate the build-up of the ice layers, tracking adsorption and desorption of the different species.  Diffusion during this time is ignored as it is expected to be minor compared to the other processes during deposition.  After deposition, the deposited sample was warmed and any gas that came off was pumped out of the experimental chamber at a constant rate.  To simulate this, we increase the temperature of the solids by 1 K per minute, with all desorbed species being removed (gas phase abundance set to zero) to simulate the pumping of the chamber.  

In order to demonstrate how the model predicts the behavior of such a mixed ice during warming and reproduces the physical effects seen in experiments, Figure 2 shows the results of a calculation  for a H$_{2}$O-CO$_{2}$ mixture of gas, at a 5:1 ratio, that was deposited at a temperature of 10 K.  Such a scenario was investigated experimentally and theoretically by \citet{fayolle11}.
  Only a thin layer of ice formed in the experiments, approximately 20 ML.  The best fit parameters for the three-phase model were found to be: $E_{\mathrm{CO_{2}}}$=2440 K, and $c_{\mathrm{CO_{2}}}$=20.5 ML.  The release of CO$_{2}$ at around 80 K is due to the relatively high binding energy of the molecule to H$_{2}$O; this is the equivalent of the release of frozen, untrapped, gases in the experiments by \citet{notesco03}.  The results presented here match well those presented in \citet{fayolle11}.

\begin{figure}[!h]
\includegraphics[width=3.5in]{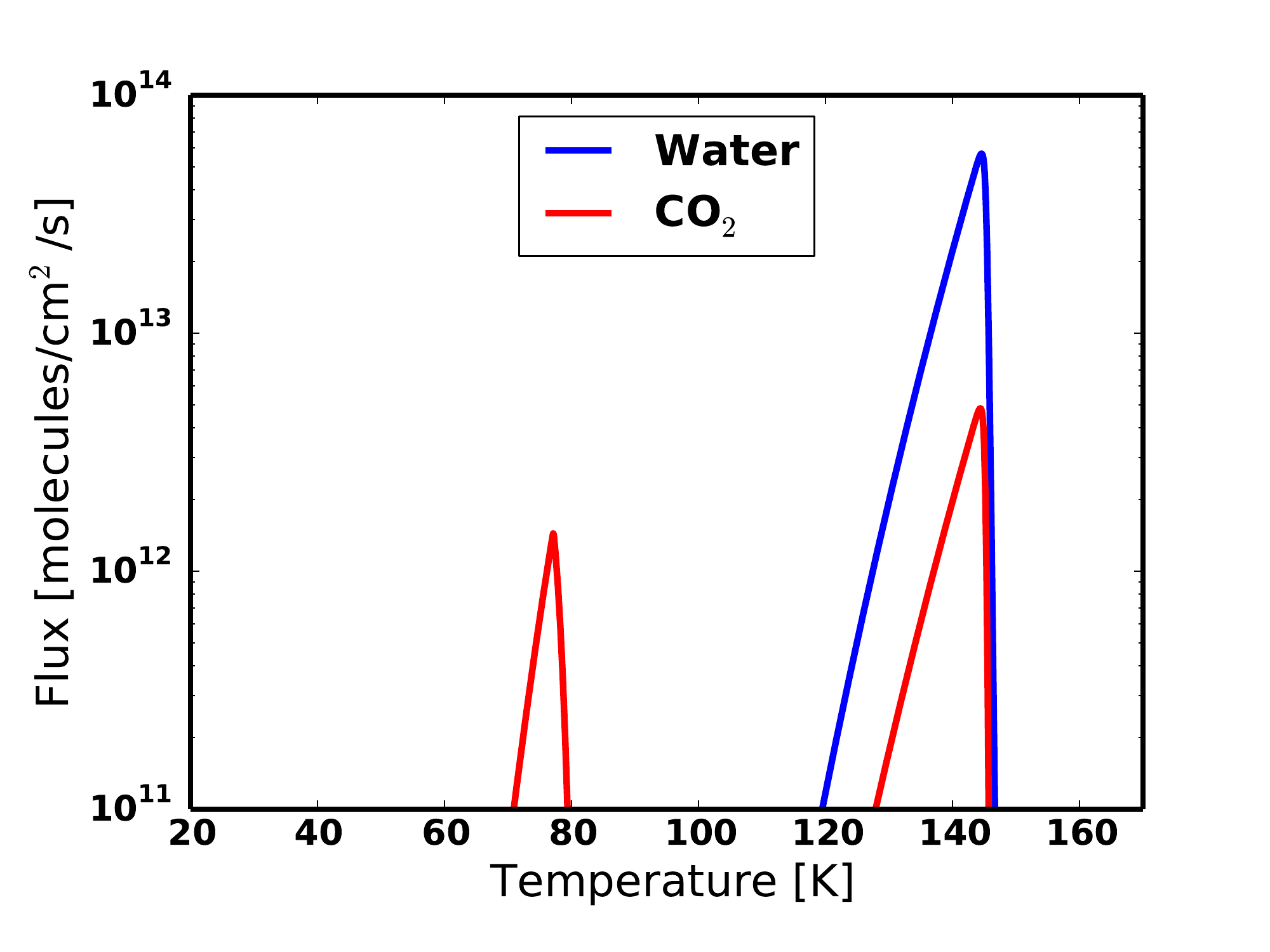}
\caption{The flux of CO$_{2}$ and H$_{2}$O coming off of the ice as it is warmed in the calculations reproducing those of \citep{fayolle11}.  The peak at $\sim$80 K arises from frozen CO$_{2}$ that desorbs directly from the surface of the ice, either as it was located there after deposition or because it diffused to there from the mantle.  As water begins to desorb from the ice at $\sim$120 K, frozen CO$_{2}$ is also found to come off the ice.  This CO$_{2}$ is assumed to have been trapped within the ice.}
\end{figure}

The relation of the swapping energy to the binding energy for CO$_{2}$ was slightly different in \citet{fayolle11}, resulting in $E_{diff}$=1520 K, or 
$E_{\mathrm{diff}}$/$E_{\mathrm{CO}_{2}} \sim $0.6, instead of 0.5 as assumed here.
In that study, however, they found that the modeling results were relatively insensitive to the value of $E_{\mathrm{diff}}$, with values ranging from 0.25-0.9$E_{\mathrm{CO_{2}}}$ matching the experimental results equally well.  Instead, it was the availability coefficient, $c_{\mathrm{CO_{2}}}$ which had a stronger control on the amount of the volatile that was trapped versus that which could be freely released via desorption.  We also find that our results do not vary significantly if we assume a different ratio of $E_{\mathrm{diff}}$/$E_{\mathrm{CO_{2}}}$, suggesting that our results will remain robust even if the relationship between these energies is more complicated than assumed here.

Turning back to Ar, we applied our model to simulate the experiments  reported in \citet{notesco03} for T=22 K and 27 K \footnote{The deposition rates and Ar/H$_{2}$O ratios were taken from the plots in that paper using datathief: \url{http://datathief.org}}.  A smaller number of experiments were also performed at 50 K, though these experiments did not see any frozen Ar being present, making them less useful in our full parameter space search here.  We return to experiments at these higher temperatures in our discussion further below.  In all cases, the gas was assumed to exist above the cold plate at the temperatures of the experiment and at
abundances such that the deposition fluxes summed to the value given by Equation (7), with $F_{\mathrm{H_{2}O}}=F_{\mathrm{Ar}}$ as the gas was meant to have a 1:1 mixture of the two species.  As the gas was flowed across the cold plate, we assumed these conditions remained fixed throughout the deposition period.
  The experiments yielded ice layers which were $\sim$0.1 $\mu$m thick, thus the time of deposition was set by $t_{dep}$=0.1/$DR$ (ranging from 1 minute to 1 week).  In each case, we varied the binding energy and availability coefficient, $E_{\mathrm{Ar}}$ and $c_{\mathrm{Ar}}$, to compare model predictions to the experimental results.  While the amount of trapped Ar is calculated for a given availability coefficient using Equation 7, the earliest stages of heating were simulated to ensure all frozen Ar was desorbed before temperatures increased much above the 40 K limit found in the experiments.

\begin{figure}
\includegraphics[width=3.5in]{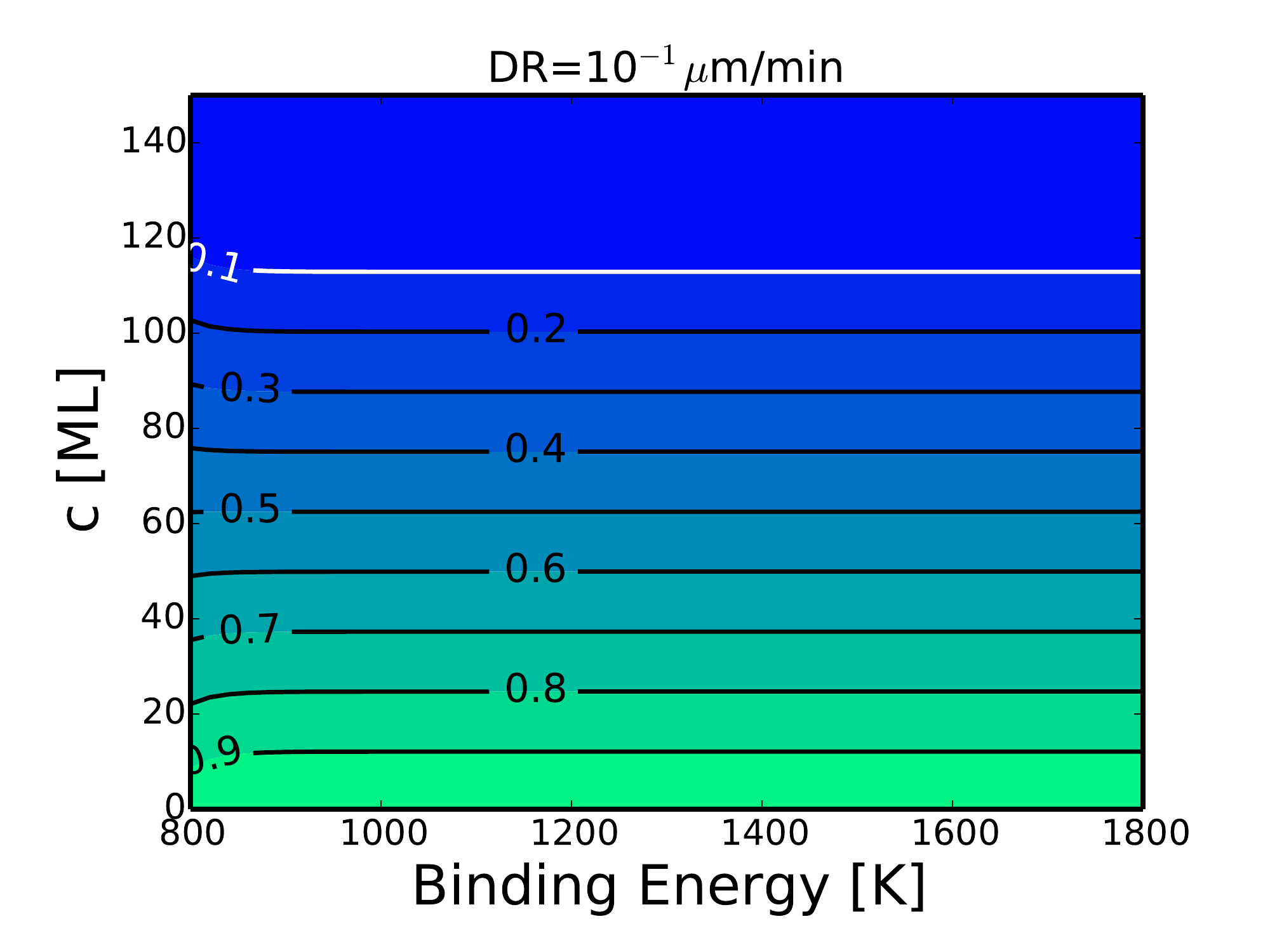}\\
\includegraphics[width=3.5in]{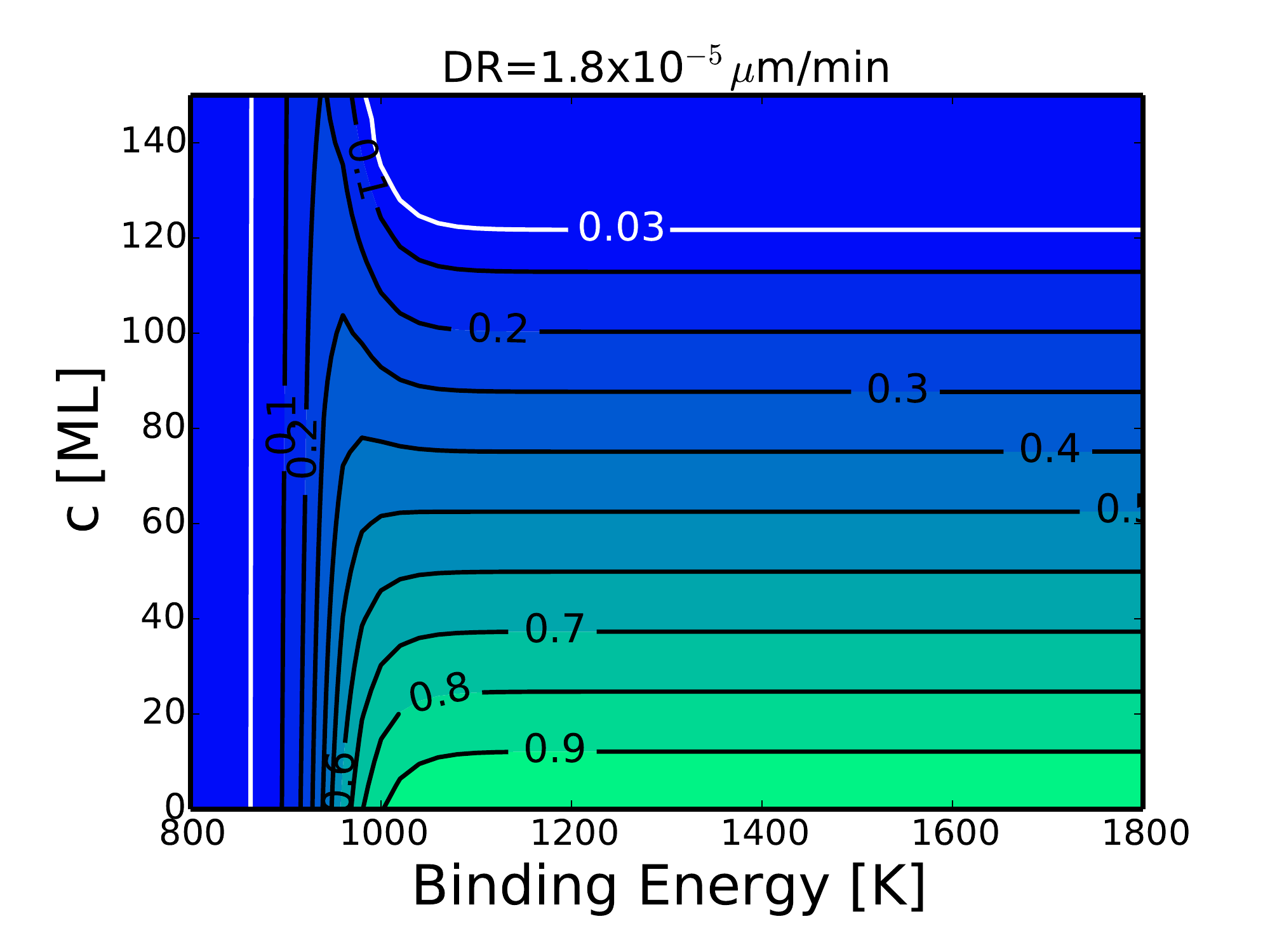}
\caption{Contour maps showing the ratio of Ar/H$_{2}$O remaining in the ice after heating for deposition temperatures of $T$=27 K model calculations, with deposition rates of 10$^{-1}$ (top) and 1.8$\times$10$^{-5}$ (bottom) $\mu$m/min, for various values of assumed Ar binding energy and availability for diffusion.  White contours indicate the value of Ar/H$_{2}$O found in the experiments.}
\end{figure}

Figure 3 shows the results of a suite of model runs for deposition rates of $DR$=10$^{-1}$ and 1.8$\times$10$^{-5}$ $\mu$m min$^{-1}$ at a temperature of 27 K, two sets of experimental conditions reported by \citet{notesco03}.  In those experiments final, trapped Ar/H$_{2}$O ratios of 0.1 and 0.03 were found within the ice respectively.  The trapped Ar/H$_{2}$O ratios in the models are shown by the contours throughout the parameter space (varying $E_{\mathrm{Ar}}$ and $c_{\mathrm{Ar}}$) that was explored.  

We see two regimes of behavior in the cases explored in Figure 3.  In the rapid deposition case, the final Ar/H$_{2}$O ratio of the ice is independent of the binding energy, $E_{\mathrm{Ar}}$, and purely set by the availability constant $c_{\mathrm{Ar}}$.  As the availability constant sets the amount of Ar that is trapped in the ice for a given mixing fraction after deposition  ($x_{\mathrm{Ar}}$), this means that the mixing fraction after deposition in this case was nearly the same for all binding energies considered here.

In the slow deposition case, we see a regime, at low binding energies, where the amount of trapped Ar becomes set purely by the binding energy and is nearly independent of the availability constant.  In these cases, we see very small values for the trapped Ar abundance (trapped Ar/H$_{2}$O in the ice $<$0.2).  These binding energies ($<$ 1000 K) yield a short residence time for Ar on the ice.  A water molecule thus has a small, finite amount of time to cover the Ar before the Ar is desorbed.  This results in only a tiny number of Ar atoms being incorporated as the ice thickens.  Due to the low concentration of Ar ($x_{\mathrm{Ar}}$) that remains in the ice, a very high amount is trapped, leaving negligible amounts to migrate and be lost during initial heating.

For the cases illustrated here, binding energies above $\sim$1000 K are high enough so that at 27 K essentially all adsorbed Ar atoms remain on the surface long enough during deposition to be buried and incorporated into the mantle.
 There is slight variation on the minimum binding energy where this occurs for the different deposition rates;  the cases with higher deposition rates allow Ar to be buried more readily as the timescale for forming monolayers (and burying adsorbed species) is lower, thus the residence time of Ar need not be as long. 
 
\begin{figure}
\includegraphics[width=3.5in]{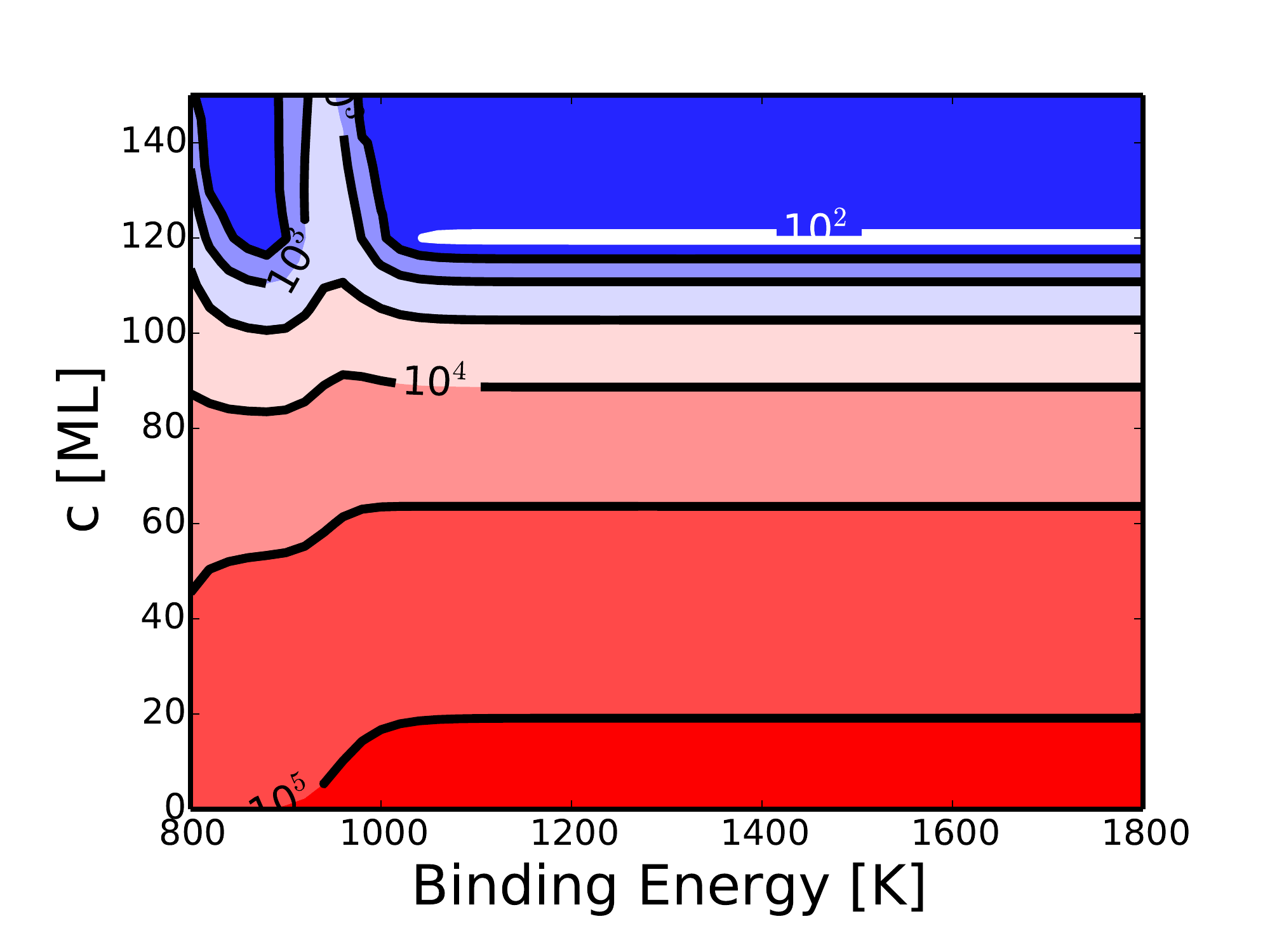}
\caption{Contours for the sum of squares of the weighted residuals, $\chi^{2}$, in the parameter space for  the T=22 K and T=27 K experimental results.  Similar structure in contour plots are seen as in Figure 3 as a result of the processes described in the text. The lowest values (best fit) are for $E_{\mathrm{Ar}} \sim$1000-1600 K and $c_{\mathrm{Ar}}$=120 ML.}
\end{figure}

Figure 4 shows the sum of the squares of the weighted residuals for our parameter exploration:
\begin{equation}
\chi^{2} = \sum \left( \frac{r_{i}^{m} - r_{i}^{e}}{\sigma_{i}^{e}} \right)^{2}
\end{equation}
where $r_{i}^{m}$ is the model prediction for the trapped Ar/H$_{2}$O ratio, $r_{i}^{e}$ is the experimentally determined ratio, and $\sigma_{i}^{e}$ is the uncertainty on the measured rate, which we take as 20\% of the experimental value as these were the variations seen in experiments when a experimental conditions were repeated \citep{notesco05}.  Here we have just applied the model fit to the T=22 K and T=27 K runs. In the T=50 K case, there is no frozen Ar, thus those cases provide little constraint on $c_{\mathrm{Ar}}$.  Further, given the experimental chamber described above, these cases could have the greatest temperature gradients, leading to greater uncertainty in the results.

 The fits indicate that  the experimental results are reproduced well by $E_{\mathrm{Ar}} >$1000 and $c_{\mathrm{Ar}}$=120 ML.  These constraints are consistent with similar efforts by  \citet{fayolle11}, who also found that a wide range of binding energies were able to reproduce the experimental results for CO$_{2}$ and CO, while only a small range of availability constants gave satisfactory fits.   Given these results, we take our best fit parameters to be $E_{\mathrm{Ar}}$=1010 K and $c_{\mathrm{Ar}}$=120 ML.   While this energy is on the low end of the well-fit range, the higher energies can be ruled out as they would lead to frozen Ar being released at temperatures much greater than 40 K, inconsistent with the experimental results.  Further, if the trapped Ar abundances reported in the experiments represent an upper limit in terms of Ar/H$_{2}$O ratios in the ice, then the fitted values for the binding energy would also represent an upper limit on this parameter, favoring a lower binding energy.  We further justify this choice of the binding energy below when considering the experimental results of \citet{yokochi12}.

 Figure 5 shows the calculated release of vapor in one of our models for a deposition temperature of 27 K and deposition rate of 10$^{-3}$ $\mu$m/min using our best fit parameters.  The general behavior shown here is the same as in the CO$_{2}$ release illustrated Figure 2 and in good agreement with the experiments described by \citet{notesco03}.  That is, frozen Ar is released up to temperatures around T$\sim$40 K, and then all other Ar remains within the solid until water begins to vaporize at T$>$120 K.  This shows that the binding energy used here is appropriate, as higher values would have led the frozen Ar desorb at too high of a temperature, while lower values would have had it desorb when it was too cool.  Further,  when models are run with deposition temperatures of 50 K, no frozen Ar is seen, again in agreement with the experiments of \citet{notesco03}.

\begin{figure}[!h]
\includegraphics[width=3.5in]{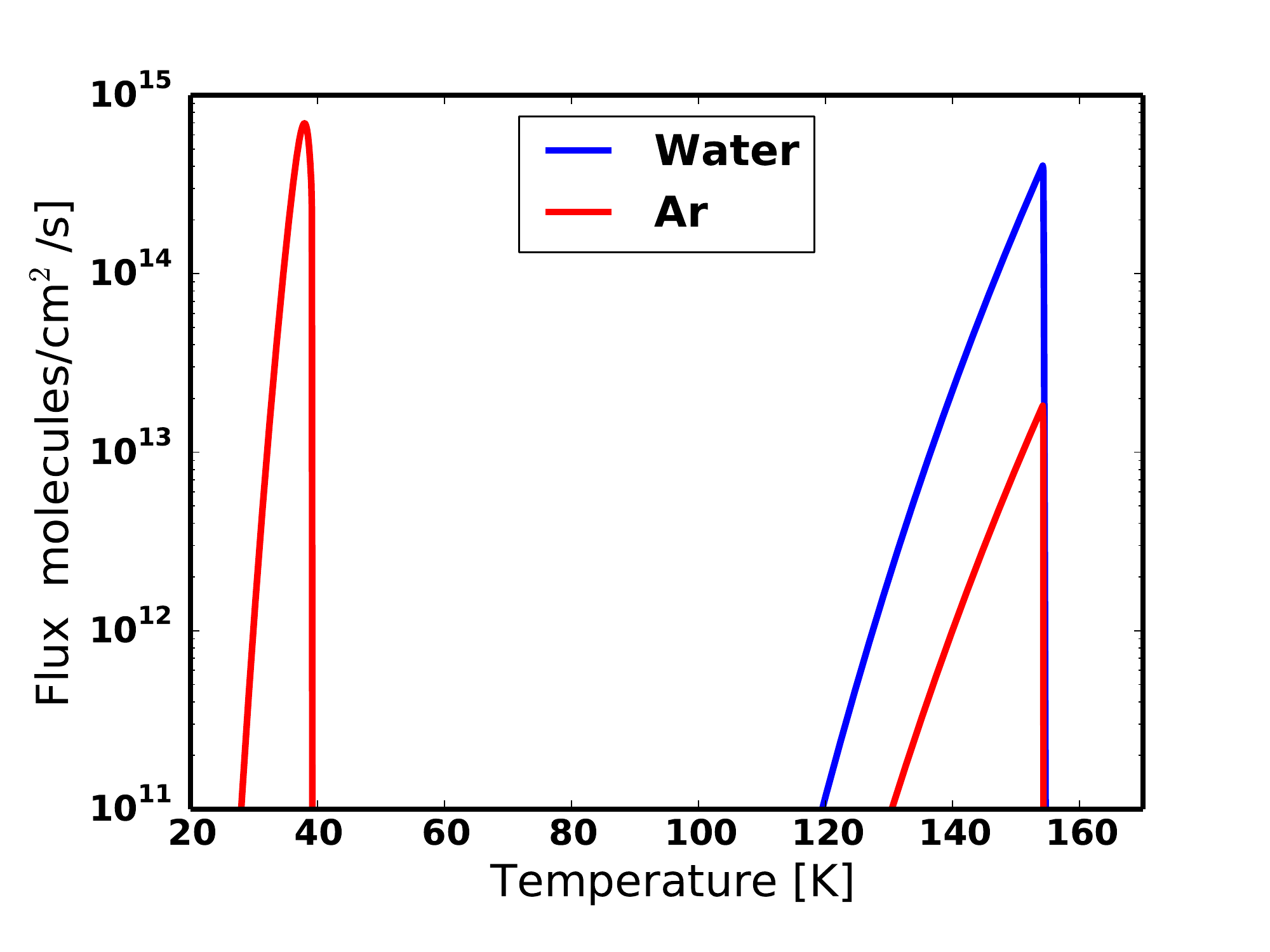}
\caption{Similar to Figure 2, but for Ar and H$_{2}$O release.  Frozen Ar is released
around $\sim$40 K, while the trapped Ar is released concurrently with the water vapor, in agreement with \citet{notesco05}.  Note that it is instantaneous flux that is plotted here as a function of temperature; while the water flux does not reach the same height as the Ar plot at lower temperatures, the integrated fluxes indicate that an equal amount of Ar and H$_{2}$O were present when warming began.  }
\end{figure}

\section{Extrapolation to Astrophysical Conditions}

The experimental studies to date have generally been done at deposition rates of 10$^{-3}$ $\mu$m/min or higher.  At 20 K, this would correspond to a water vapor density of n$_{\mathrm{H_{2}O}} \sim$10$^{10}$ cm$^{-3}$.  Water is typically present in a gas of solar compostion with a ratio of $n_{\mathrm{H_{2}O}}$/$n_{\mathrm{H_{2}}}$=5$\times$10$^{-4}$ \citep{lodders03,cleeves14}, meaning the experimental fluxes correspond to environments with n$_\mathrm{H_{2}}$ of 2$\times$10$^{13}$ cm$^{-3}$ or higher.  Such densities are generally expected towards the inner regions ($<$5 AU) of a protoplanetary disk \citep[densities of $\sim$10$^{-10}$ g cm$^{-3}$ or pressures of $\sim$5$\times$10$^{-8}$ bars, e.g.][]{ciesla_dull10,bergin_pdchem11}.  However, amorphous ice is much more likely to form in the very outer regions of the solar nebula, possibly above the disk midplane, or in the natal molecular cloud from which the solar system formed \citep{kouchi94,ciesla14}.  In these cases, hydrogen number densities likely were much less, possibly as low as  $n_{\mathrm{H_{2}}}$=10$^{3}$-10$^{10}$ cm$^{-3}$ \citep{bergin_pdchem11}, implying deposition fluxes that were as much as 10-12 orders of magnitude lower than those used in experiments.

Taking the best fit parameters from above, we can apply the three-phase model to examine how much Ar would be trapped at the much lower deposition rates expected in these astrophysical environments. We assume a gas of solar composition \citep[Ar/H$_{2}$=5$\times$10$^{-6}$][]{asplund09}, which gives gas phase ratios Ar/H$_{2}$O=0.01, assuming all the water is initially present as a vapor.  We consider temperatures of 10, 20, 30, and 40 K and deposition rates ranging from the 10$^{-1}$ $\mu$m/min used in the experiments, down to 10$^{-16}$ $\mu$m/min.  The extremely slow deposition rates would obviously require time periods longer than the age of the universe to produce ice layers of the thicknesses modeled here; these are not meant to be realistic scenarios, but rather show how the trapping behavior, even at low temperatures, varies with deposition flux in order to develop an intuitive understanding of key parameters.   In all cases, the same approach was followed as before: ice was deposited to form a layer measuring 0.1 $\mu$m thick at constant deposition rates and temperatures, with all gases replenished as in the experiments.  While this would not be the case in reality if a particle stayed dynamically bound to the same gas in a given environment, this was done to make comparisons to the experiments easier; we return to this issue further below.  

Figure 6 shows the results of these calculations, where two regimes of trapping are readily seen.  The first is a \emph{burial} regime, where the Ar/H$_{2}$O ratio of the ice is roughly comparable to that of the gas ($\sim$0.01).  This is seen at low temperatures and high deposition rates.  The second is a \emph{equilibrium} regime, where the trapped Ar/H$_{2}$O ratio is proportional to the deposition rate.  This is seen at higher temperatures and low deposition rates.  The transition between the two trapping regimes represents a shift in the likelihood of an adsorbed Ar atom being trapped by water molecules during deposition.  This shift  occurs when the timescale for a monolayer of water to be deposited becomes comparable to, or exceeds, the residence time for an Ar atom on the surface of the ice.  That is, the typical amount of time that an Ar atom would spend on the surface after adsorption would be
 $t_{res}$=$\nu_{\mathrm{Ar}}^{-1} \exp(\frac{E_\mathrm{Ar}}{T})$, while the timescale for a monolayer of water ice to form is $t_{ML}$=$N_{s}$/$F_{\mathrm{H_{2}O}}$.    In all cases, Ar is continuously adsorbed onto the surface of the ice.  For cases when $t_{res} > t_{ML}$, 
thse Ar atoms sit on the surface long enough to be covered by layers of water ice, meaning that their abundance is set only by how quickly they are delivered to the ice.
  When $t_{res} < t_{ML}$, Ar atoms are relatively transient, and thus many are able to desorb before being covered with water.  This allows the total amount of Ar on the surface to be set by the relative rates of adsorption and desorption, resulting in some equilibrium coverage for this species.

This equilibrium trapping was observed in experiments by \citet{yokochi12}, where the amount of Ar that was trapped in amorphous water ice varied with the pressure of Ar during ice deposition.   These authors proposed that the pressure dependence seen in their experiments (Ar/H$_{2}$O ratio was  proportional to the partial pressure of Ar in the experiments) was due to this equilibrium effect, where the amount of Ar on the ice was able to rapidly adjust before being buried by a layer of trapping water.    As these experiments were performed at $\sim$77 K, the high temperatures would imply very short residence times of any guest species, putting them squarely in the equilibrium regime.  Within the framework here, the equilibrium abundance of Ar on the surface of the grain would be set by setting the desorption and adsorption fluxes onto the substrate \citep{bergin_pdchem11}:
\begin{equation}
\frac{1}{4} n_{\mathrm{Ar}}^{g} v_{th}^{Ar} = n_{\mathrm{Ar}}^{s} \nu \exp{\left(-\frac{E_{\mathrm{Ar}}}{T}\right)}
\end{equation}
Writing $n_{\mathrm{Ar}}^{g}$ as $P_{\mathrm{Ar}}$/$kT$, we see that the abundances of Ar on the surface will be proportional to the partial pressure of Ar.  Following this, we can write the Ar/H$_{2}$O ratio of the ice to be:
\begin{equation}
\frac{n_{\mathrm{Ar}}^{s}}{n_{\mathrm{H_{2}O}}^{s}} =
\frac{v_{th}^{\mathrm{Ar}}}{4kT n_{\mathrm{H_{2}O}}^{s}}  \nu^{-1} \exp{\left(\frac{E_{\mathrm{Ar}}}{T}\right)}P_{Ar}
\end{equation}
 \citet{yokochi12} found that the  Ar/H$_{2}$O ratio in the ice was given by 2.34$\times$10$^{-4} P_{\mathrm{Ar}}$, if the pressure is given in $\mu$bars. Taking that value as the proportionality factor in Equation (11), and assuming that in the equilibrium stage that water dominates the surface (n$_{\mathrm{H_{2}O}}^{s} \sim$ 10$^{15}$ cm$^{-2}$) allows us to estimate $E_{\mathrm{Ar}}$ = 1010 K, justifying our choice above.  Further, we also modeled cases where the Ar/H$_{2}$O ratio in the gas was varied for those cases which were found to be in the \emph{equilibrium} regime.  We found that the final Ar/H$_{2}$O ratio in the ice was independent of this value in the gas provided $t_{res} < t_{ML}$, again in agreement with the experimental findings of \citet{yokochi12} \footnote{As discussed by \citet{yokochi12}, their experiments generally found lower trapping efficiency than those in \citet{barnun88}.  This may be due to different experimental approaches as \citet{yokochi12} considered trapping in a closed system, such that Ar could only freeze-out with water.  As discussed above,  the numbers from the earlier experiments should be considered upper limits on the trapping efficiency rather than absolute values, suggesting that the two results may be in agreement with one another.}

 \citet{smith16} also reported results from an experimental study of the desorption of Ar from amorphous ice.  Their experiments were not focused on the trapping of Ar during ice deposition; rather they deposited Ar at a temperature of 25 K on the surface of already formed amorphous ice, then examined when Ar desorbed off of the surface upon heating.  They found that the desorption relationship followed a similar formula as that used here (Eq. 2), but with a pre-exponential factor of $\nu$=6.2$\times$10$^{11}$ s$^{-1}$ and $E_{\mathrm{Ar}}$=870 K.  Using these values and following the arguments from above, would lead to a predicted transition between equilibrium and burial trapping at higher deposition rates for the temperatures of interest considered here.  Specifically, the residence times using the \citet{smith16} numbers would be shorter by a factor of $\sim$20 at 40 K, and $\sim$700 at 20 K.  At 77 K, the  residence time predicting is changes by a factor of just $\sim$4, leaving the \citet{yokochi12} experiments still squarely in the Equilibrium regime.
 
Diffusion only plays a minor role in the trapping of Ar in these cases.  This is due to the relatively low abundance of Ar in the deposited ice ($x_{\mathrm{Ar}}^{ini} \le 0.01$), which according to Equation 6, means most ($>$99\%) of the Ar in the mantle is prevented from exchanging with the surface.    Diffusion would only become important in cases where Ar/H$_{2}$O ratios were much higher than in a gas of solar composition.   This is true even for different values of $\beta$ in Equation 6, and is consistent with experimental results suggesting that importance of diffusion decreases for high ratios of water to the trapped volatile.

\begin{figure}[!h]
\includegraphics[width=3.5in]{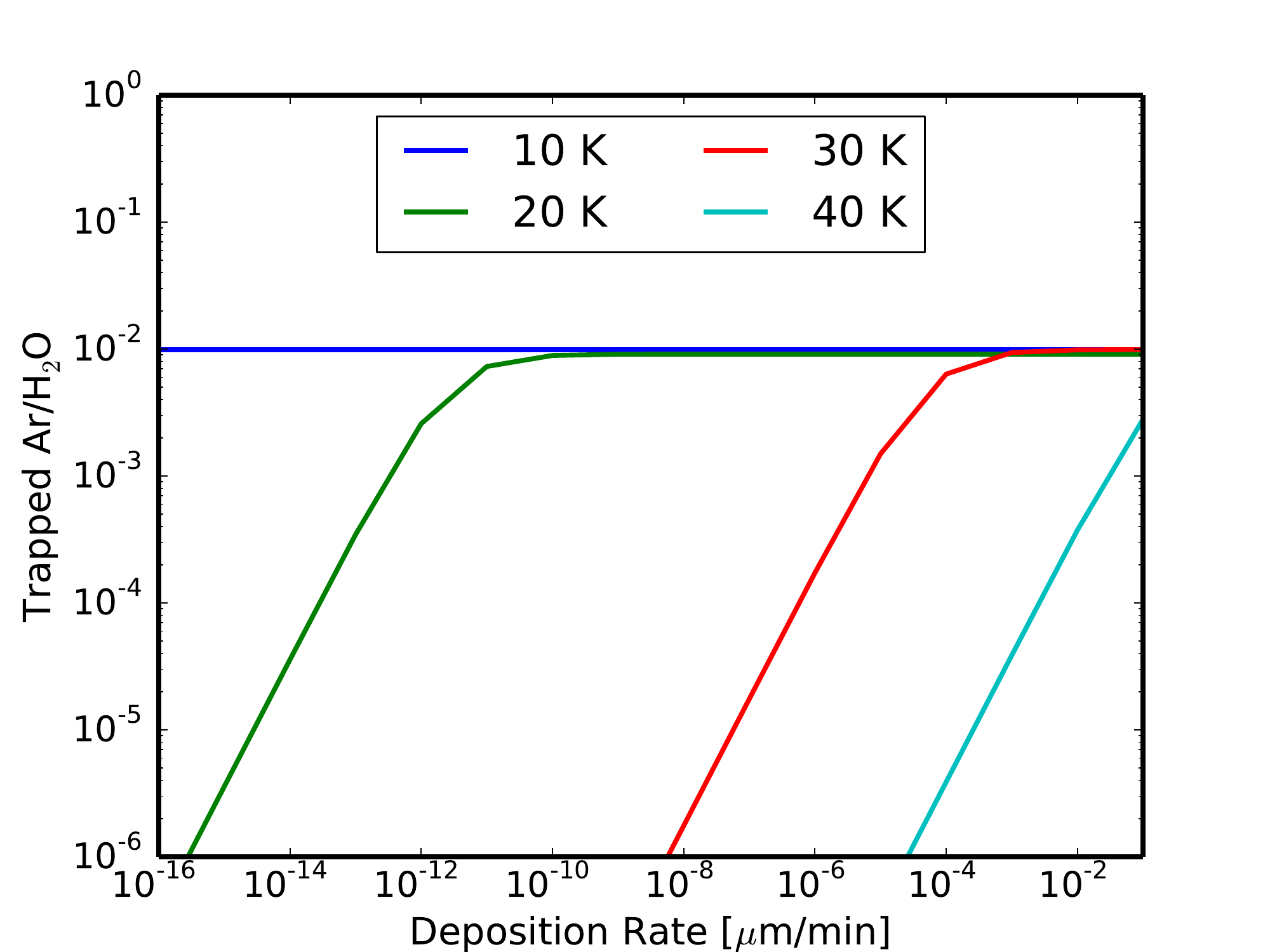}
\caption{The predicted trapped Ar/H$_{2}$O ratio of ices formed in a gas of solar composition at various temperatures and deposition rates.  The \emph{Burial Regime} is reached when the Ar/H$_{2}$O ratio in the ice mirrors that of the gas and is independent of deposition rate.  The \emph{Equilibrium Regime} is reached when the abundance of Ar is a function of deposition rate.  The transition between these regimes occurs approximately where the timescale for desorption of an Ar atom is comparable to the timescale for a monolayer of water ice to form over that Ar atom.}
\end{figure}

The model used here necessarily simplifies some of the processes that are likely at work when amorphous water ice forms and traps guest species.  For example, it was assumed that the structure of the water ice was independent of the formation temperature.  \citet{notesco03} discussed how the surface area of water varies depending on the conditions at which it is deposited. This is likely related to the ability of water molecules to arrange in the binding sites available in the layers provided by a given surface, and as such the morphology of the ice likely varies with the temperature of formation.  This is important as nanopores in the ice may provide sites where adsorbed gases may be more strongly adsorbed (have a higher binding energy) than other sites in the ice; this effect has been noted for CO in amorphous water ice \citep{karssmeijer14}.  \citet{smith16} also discussed how given the structure of amorphous ice, there exists a distribution of binding site energies, with the derived values describing desorption representing the most probable value in that distribution.   Thus, binding energies are likely not constant, and the available surface area and density of surface sites may vary with deposition temperature.  These effects should be looked at in order to better understand the trapping ability of ices that form at very cold environments. \citet{visser09} considered a distribution of binding energies for CO while investigating its retention during molecular cloud collapse and protoplanetary disk formation, and documenting this distribution in detail for various species will be important.

  Further, we assumed that the deposition flux was constant throughout deposition; in real astrophysical environments, the gas phase abundance would decrease as species adsorbed onto solids, leading to changes in deposition rates with time.  As water molecules have a lower mass than Ar atoms, the rate of depletion of water vapor would be  faster.  That means that the Ar composition would not be uniform in the mantle, and the Ar/H$_{2}$O ratio would increase as it moved closer to the surface.  This could mean greater fractions of Ar would be lost via desorption during warm up, as migration would be more important in regions of higher Ar/H$_{2}$O ratios, though this issue should be quantitatively examined by future experiments.

While the model used here was simplified, it has successfully reproduced many of the features and relations observed in amorphous ice trapping experiments.  Thus this framework is useful for understanding how species may be trapped in amorphous ice.  Future experimental studies should be interpreted in this context, with the three-phase model allowing extrapolation from the laboratory conditions to astrophysical environments.

\section{Summary and Discussion}

Here we have applied the three-phase model to determine which parameters allow us to reproduce the experimental trapping results of \citet{notesco03} and applied that model to conditions that are more similar to those expected during the stages of planet formation.  We have identified two different trapping regimes: \emph{burial} where the composition of the ice reflects the composition of the gas during deposition, and \emph{equilibrium} where the amount of trapped species scales with the density of the gas being trapped.  Both regimes have been identified in experimental studies; here we have developed a model to identify under what conditions each would occur.  Burial trapping occurs under conditions that allow for  rapid ice deposition and when temperatures are very low; this allows gases to reside on the surface for long enough that they can be covered by another layer of deposited ice.    Equilibrium trapping occurs when temperatures are higher or the rate of ice deposition is slow, allowing for the the amount of Ar on the surface to be set by both adsorption and desorption.   

If trapping of noble gases in amorphous ice is necessary for explaining the features we see in planetary atmospheres and small bodies, then the results described here help to constrain the formation conditions for those ices.  
Forming solar-composition solids that include noble gases as envisioned by \citet{owen99}, requires conditions that result in burial trapping. For the temperatures of 10, 20, 30, and 40 K, the transition between the burial and equilibrium trapping  (when  $t_{res} = t_{ML}$) occurs at deposition rates of $\sim$3$\times$10$^{-34}$, 3$\times$10$^{-12}$, 6$\times$10$^{-5}$, and 0.3 $\mu$m min$^{-1}$, respectively (approximately where the transition from sloped to horizontal  line occurs for the  curves in Figure 6).  These deposition rates would  correspond to environments with hydrogen densities of $n_{\mathrm{H_{2}}} \sim$10$^{-18}$,  10$^{4}$, 10$^{11}$, and 10$^{15}$ cm$^{-3}$ respectively, assuming all water is present as a vapor at a mixing ratio of H$_{2}$O/H$_{2}$=5$\times$10$^{-4}$.  That is, burial trapping would occur provided environments were more dense than these critical values for the respective temperatures.  At 10 and 20 K, the critical values are generally so low that molecular cloud or protoplanetary disk environments exceed them, implying that burial trapping would occur at these temperatures.  

In the discussion thus far, we have assumed that water is present in the vapor when deposition occurs.  This generally not expected for the low temperatures where trapping is likely to occur \citep[e.g.][]{frayschmitt09}.  As shown here, water will largely exist at a solid at temperatures $<$100 K and is expected to be frozen out at these temperatures even at the pressures found in the interstellar medium \citep{sandfordall90,fraser01}.  Thus providing the water that will trap the noble gases at these temperatures requires some event or exterior input of energy that would lead to water molecules being liberated and then freezing-out at the fluxes defined above.  If some heating event were to occur (e.g., shock wave, impact plume), that vaporized the water, the cooling timescale would have to be faster than the water freeze-out timescale in order for enough water to be present to then freeze-out at the relevant temperatures.  The timescale for freeze-out of water is \citep{bergin_pdchem11}:
\begin{equation}
t_{fo} = 2 \times 10^{4} \left( \frac{5 \times10^{4} \mathrm{~cm}^{-3}}{n_{\mathrm{H_{2}}}} \right) \left( \frac{20 \mathrm{~K}}{T} \right) \mathrm{~ yrs}
\end{equation}
assuming that the solids would be present at ISM abundances and sizes.  Grain growth or depletion of solids would increase the freeze-out timescales, though may also limit how much water is vaporized. The inverse dependence of the freeze-out timescale on gas density  favors low-density environments for trapping (as it would make it easier to satisfy $t_{cool} < t_{fo}$).     

An alternative method by which water may be liberated into the vapor at these low temperatures is through photodesorption, where UV photons provide the energy for molecules to be lost from the surface \citep{westley95,oberg09b}.  This effect is believed to be responsible for the cold water vapor seen in the outer regions of disks like TW Hydra \citep{hogerheijde11}.  \citet{ciesla14} showed that the movement of grains into the surface regions of the disk via turbulent diffusion could expose them to sufficient UV to lose most, if not all of the water on their surfaces, only to have the water freeze-out at very high fluxes as the gas and vapor diffused toward the midplane again.  The corresponding fluxes of water during freeze-out would have been $\sim$10$^{5}$-10$^{9}$ molecules cm$^{-2}$ s$^{-1}$.  This yields $t_{ML}$ of 10$^{6}$-10$^{10}$ s, suggesting burial trapping can occur provided temperatures are $<$25 K (when $t_{res} > t_{ML}$).

\citet{monga15} also invoked photodesorbed water as a means of producing amorphous ice and trapping noble gases to eventually enhance the abundance in Jupiter.  In that study, solids had grown and settled to the midplane, reducing the available surface area on which water molecules could freeze-out.  As a result, water molecules would diffuse downward and outward from where they were photodesorbed for timescales of $\sim$10$^{3}$ years or longer, resulting in $\sim$0.1$M_{\oplus}$ of water vapor in an annulus ranging from 30 to 50 AU.  Assuming the gas is uniformly distributed over a height of 2 scale-heights (one on each side of the disk midplane with $H \sim 0.05r$), this would imply a water volume density of $n_{\mathrm{H_{2}O}} \sim$10$^{5}$ cm$^{-3}$, suggesting a freeze-out flux of 10$^{9}$ molecules cm$^{-2}$ s$^{-1}$.  Again, this would mean burial trapping will occur in those regions only where temperatures $<$25 K.  

These two scenarios only consider UV photodesorption as a source of cold water vapor.  Sputtering by cosmic rays can also liberate water from ices \citep[e.g.][]{dartois15}, though the importance of this effect will depend on the cosmic ray flux.  It is possible that stellar winds from the young Sun prevented significant penetration of cosmic rays into the solar nebula \citep{cleeves13}, which would limit this effect.  However, if trapping occurs within dense molecular clouds, this may be an important source of water vapor.

The exact amount of water that is available in the gas phase is also important in determining how much Ar will be trapped as amorphous ice forms.
The discussions thus far has assumed that water is present at its solar abundance relative to Ar when trapping occurs.  However, it is possible that only a portion of water will be removed in whatever process produces the cold water vapor, or that some fraction will freeze-out before conditions reach those where noble gases can be effectively trapped.  If only a portion of the water is available in the gas, this would alter the ratio of Ar/H$_{2}$O in the vapor, and thus the resulting mix in the ices that are deposited on the surfaces of the grains.  As discussed above, the solar ratio of Ar/H$_{2}$O$\sim$0.01 means that low Ar concentration would lead to very large fractions, nearly all of it, being trapped in the water ice.  If only $\sim$1\% of the water is liberated into the gas, however, then the Ar/H$_{2}$O ratio in the vapor would be 1:1, similar to the experiments that were carried out by \citet{notesco03} and \citet{notesco05}.  In the context of the model presented here, this would increase the abundance of Ar in the outermost ice layers, increasing the likelihood of Ar diffusing and desorbing from the surface of the grains upon warming, reducing the efficiency of this process.  However, experiments have also indicated that once a
 trapped species exceeds $\sim$25-30\% of the H$_{2}$O abundance, the H-bonding network is effectively disrupted and the trapped species can leave far more easily \citep{sandfordall88}.  Thus burial trapping is unlikely under these conditions; it is much more likely to occur in environments or conditions where large amounts of water are vaporized.  In the models of \citet{ciesla14}, large freeze-out fluxes tended to occur when large amounts of water were vaporized; these conditions are ideal for burial trapping, but occur less frequently as they require movement to higher altitudes above the disk midplane.  In \citet{monga15}, high Ar/H$_{2}$O abundances are expected at lower heliocentric distances, which reduce the fraction of Ar that is trapped instead of frozen.  This suggests that efficient trapping would have to be limited to the very outer regions of the zone they had envisioned.

Thus, the best environments for trapping noble gases in amorphous ice will be those where temperatures are $<$25 K and water is liberated into the gas such that it is at least $\sim$10$\times$ more abundant than those species.  The very outer regions of protoplanetary disks and molecular clouds, where photodesorption or localized heating events occur, would be the best candidates.  However, even if trapping does occur in these locations, what gets delivered to comets or planets will ultimately depend on what happens to the icy particle after trapping occurs.  Within the solar nebula, individual grains are likely to see a wide range of physical environments once added to the disk as part of the infall and early evolution \citep[e.g.][]{visser09} or as they are subjected to dynamic processes within the disk \citep{cieslasandford12}.  When amorphous H$_{2}$O-rich ices are warmed they can go through several intermediate phase changes before the H$_{2}$O sublimes.  This includes (i) a change between two different amorphous states  (ii) a transition from amorphous ice to cubic crystalline ice (at ~120K), and, (iii) just as the ice is subliming, a partial transition to hexagonal ice (at ~150K) \citep[e.g.][]{barnun88,sandfordall88,blake91}.  This can result in the expulsion of some trapped volatiles \citep{sandfordall88,collings03}, though it is possible for some to be retained, but in the form of clathrates instead of amorphous ice \citep{blake91}.  Only a small amount of volatiles are lost at any given time \citep{viti04}, but the details of transport and volatile retention will be the focus of future work.

\acknowledgments
The authors are grateful for discussions with Ted Bergin, Edith Fayolle, and Karin \"Oberg.  We also thank the referees for the very helpful suggestions that led to improvements to the manuscript.  This work was supported by NASA grants NNX14AG97G and NNX14AQ17G.  S.K. acknowledges support from NASA through Hubble Fellowship grant HST-HF2-51394 awarded by the Space Telescope Science Institute, which is operated by the Association of Universities for Research in Astronomy, Inc., for NASA, under contract NAS5-26555.  This material is also based upon work supported by the National Aeronautics and Space Administration under Agreement NNX15AD94G for the program ``Earths in Other Solar Systems.'' The results reported herein benefited from collaborations and/or information exchange within NASA's Nexus for Exoplanet System Science (NExSS) research coordination network sponsored by NASA's Science Mission Directorate

\bibliographystyle{apj}
=



\end{document}